\documentclass{article}
\usepackage{ijcai25}

\usepackage{times}
\usepackage{soul}
\usepackage{url}
\usepackage[hidelinks]{hyperref}
\usepackage[utf8]{inputenc}
\usepackage[small]{caption}
\usepackage{graphicx}
\usepackage{amsmath}
\usepackage{amsthm}
\usepackage{booktabs}
\usepackage{algorithm}
\usepackage{algorithmic}
\usepackage[switch]{lineno}
\usepackage{enumitem}
\usepackage{multirow}
\usepackage{array}
\usepackage{float}
\usepackage{caption} 
\usepackage{subcaption}
\usepackage{arydshln}
\usepackage{wasysym}
\usepackage{amsfonts,amssymb} 
\usepackage{mathrsfs}

\title{MIND-EEG: Multi-granularity Integration Network with Discrete Codebook for EEG-based Emotion Recognition}

 \author{
 Yuzhe Zhang$^{\#1}$\and
 Chengxi Xie$^{\#2}$\and
 Huan Liu$^{*1}$\and
 Yuhan Shi$^3$\and
 Dalin Zhang$^4$\\
 \affiliations
 $^1$School of Computer Science and Technology, Xi'an Jiaotong University\\
 $^2$Joint School of Design and Innovation, Xi'an Jiaotong University\\
 $^3$School of Information Science and Engineering, Lanzhou University\\
 $^4$Space Information Research Institute, Hangzhou Dianzi University\\
 \emails
 zhangyuzhe@stu.xjtu.edu.cn,
 syeyedao@gmail.com,
 huanliu@xjtu.edu.cn,
 220220943121@lzu.edu.cn,
 dalinzhang@ieee.org
 \thanks{\textsuperscript{\#}Equal contribution.}
 \thanks{\textsuperscript{*}Corresponding author.}
 }

\begin{document}

\maketitle

\begin{abstract}
        Emotion recognition using electroencephalogram (EEG) signals has broad potential across various domains. EEG signals have ability to capture rich spatial information related to brain activity, yet effectively modeling and utilizing these spatial relationships remains a challenge. Existing methods struggle with simplistic spatial structure modeling, failing to capture complex node interactions, and lack generalizable spatial connection representations, failing to balance the dynamic nature of brain networks with the need for discriminative and generalizable features. To address these challenges, we propose the Multi-granularity Integration Network with Discrete Codebook for EEG-based Emotion Recognition (MIND-EEG). The framework employs a multi-granularity approach, integrating global and regional spatial information through a Global State Encoder, an Intra-Regional Functionality Encoder, and an Inter-Regional Interaction Encoder to comprehensively model brain activity. Additionally, we introduce a discrete codebook mechanism for constructing network structures via vector quantization, ensuring compact and meaningful brain network representations while mitigating over-smoothing and enhancing model generalization. The proposed framework effectively captures the dynamic and diverse nature of EEG signals, enabling robust emotion recognition. Extensive comparisons and analyses demonstrate the effectiveness of MIND-EEG, and the source code is publicly available at \url{https://anonymous.4open.science/r/MIND_EEG}.
\end{abstract}

\section{Introduction}

Emotion recognition is a critical research area with diverse applications such as human-computer interaction, healthcare, and personalized learning~\cite{li2018novel,personality,lou2024}. While behavioral signals like facial expressions~\cite{facial} are easy to recognize for emotions, they are susceptible to intentional manipulation. In contrast, electroencephalography (EEG) signals directly measure brain activities and cannot be consciously controlled, offering high reliability. Moreover, EEG signals have high temporal resolution and the ability to capture both conscious and subconscious emotions~\cite{EEG}, making them particularly effective for emotion recognition~\cite{zhang2022eeg}.

\begin{figure}
    \centering
        \includegraphics[width=\columnwidth]{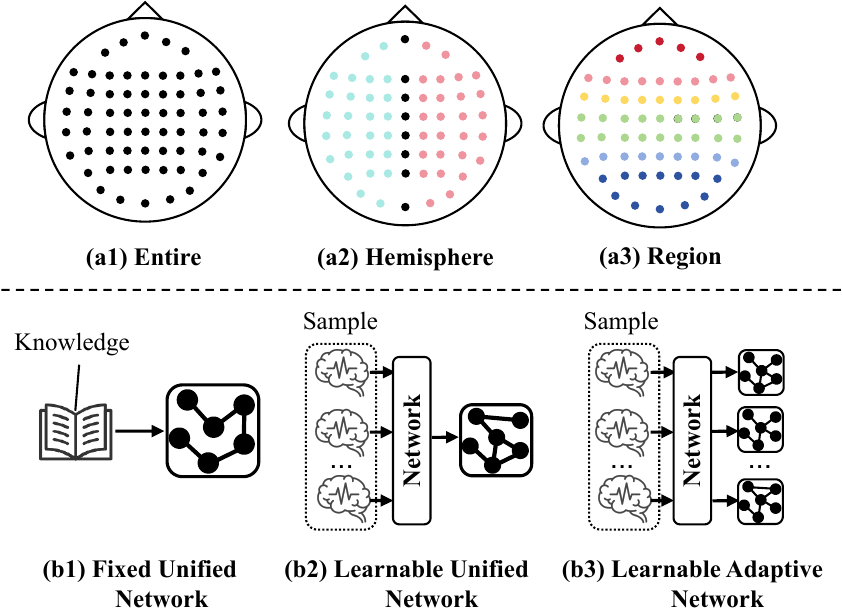}
        \caption{Two key questions for utilizing spatial information in EEG signals.}
    \label{motivation}
\end{figure}

EEG signals are inherently multivariate time series data, encompassing both spatial and temporal information. A critical research focus lies in effectively extracting spatial information, which can be approached through the following two key questions:


\textbf{Q1:} \textit{How to model the spatial structure of EEG nodes?}
As illustrated in the upper part of Figure \ref{motivation}, EEG nodes are distributed across the scalp with specific spatial arrangements but lack explicit structural relationships. Modeling these structural relationships is tantamount to modeling brain functionalities, enabling the capture of interactions and dependencies among different nodes to better recognize emotional states. Some studies approach this by treating the brain as a unified whole\cite{RGNN} (Figure\ref{motivation}(a1)), while others focus on the anatomical and functional differences between the left and right hemispheres, modeling these two parts independently\cite{BiHDM} (Figure \ref{motivation}(a2)). Additionally, some methods divide the brain into distinct regions based on functionalities, creating a multi-region structural model \cite{R2G-STNN,LGGNet} (Figure~\ref{motivation}(a3)). However, given the complexity of neural processes, relying on single structural modeling is insufficient for fully capturing the spatial features of EEG data. This underscores the necessity of a more comprehensive and integrative approach to modeling the spatial structure of EEG nodes.

\textbf{Q2:} \textit{How to represent the spatial connections between EEG nodes?}
As depicted in the lower part of Figure\ref{motivation}, existing studies can be broadly classified into three categories. Some methods construct a fixed, unified representation for all samples based on established neuroscience knowledge\cite{SGA-LSTM} (Figure~\ref{motivation}(b1)). Others adopt a data-driven approach, creating a learnable unified representation for all samples\cite{DGCNN} (Figure~\ref{motivation}(b2)). However, both approaches rely on a single unified network structure for all samples, which overlooks the dynamic and variable nature of brain network structures.
To address this limitation, certain methods adaptively learn distinct representations for each sample using neural networks~\cite{V-IAG} (Figure~\ref{motivation}(b3)). While promising, these approaches often encounter challenges such as overfitting and over-smoothing, particularly when working with small, augmented EEG datasets. In these scenarios, the neural network may compress distinct representations into a narrow region of the latent space, reducing its capacity to capture meaningful and discriminative features.

To address the above two questions, we propose the Multi-granularity Integration Network with Discrete Codebook for EEG-based Emotion Recognition (MIND-EEG). The proposed framework models the spatial structure from multiple granularities. Specifically, it has a Global State Encoder for modeling the global granularity, an Intra-Regional Functionality Encoder for modeling the regional granularity, and an Inter-Regional Interaction Encoder for modeling the granularity of inter-region interactions. 
Moreover, for each encoder, we propose an innovative approach using a codebook that stores discrete vector quantized representations of the spatial connections. This method assigns each sample a more compact, versatile, and meaningful spatial connection representation from the codebook, preventing embeddings of similar samples from converging to overly similar values, thus effectively mitigating the over-smoothing issue. Moreover, discrete embeddings are better suited for capturing inter-class differences, and maintaining clear distinctions between different categories in the latent space, thereby enhancing the model's recognition ability and generalization performance.

We conducted experiments under both subject-dependent and subject-independent scenarios on three benchmarking datasets, namely SEED-IV~\cite{SEED-IV}, MPED~\cite{MPED}, and SEED-V~\cite{SEED-V}. Our model achieved superior performance under both scenarios, significantly improving the emotion recognition accuracy compared to state-of-the-art models. 
Ablation studies demonstrate the effectiveness and robustness of each proposed component. 
Statistical studies on the codebook further validate the effectiveness of this design. The main contributions of this work are summarized as follows:

1) We propose the Multi-granularity Integration Network with Discrete Codebook for EEG-based Emotion Recognition (MIND-EEG), which performs multi-granularity modeling and integration of spatial information in EEG signals for emotion recognition.

2) We innovatively introduce a codebook that stores discrete representations in the construction of the spatial connections, to perform vector quantization on the network structures at each granularity for each sample. This ensures the diversity of the network structures and enhances the model's ability to extract class-related information.

3) We conducted extensive comparative experiments on three public datasets and under two experimental scenarios. Our model achieved significant improvements in the subject-dependent scenario and delivered competitive results in the subject-independent scenario. The source code is publicly available at \url{https://anonymous.4open.science/r/MIND_EEG}.

\begin{figure*}
    \centering
        \includegraphics[width=1\linewidth]{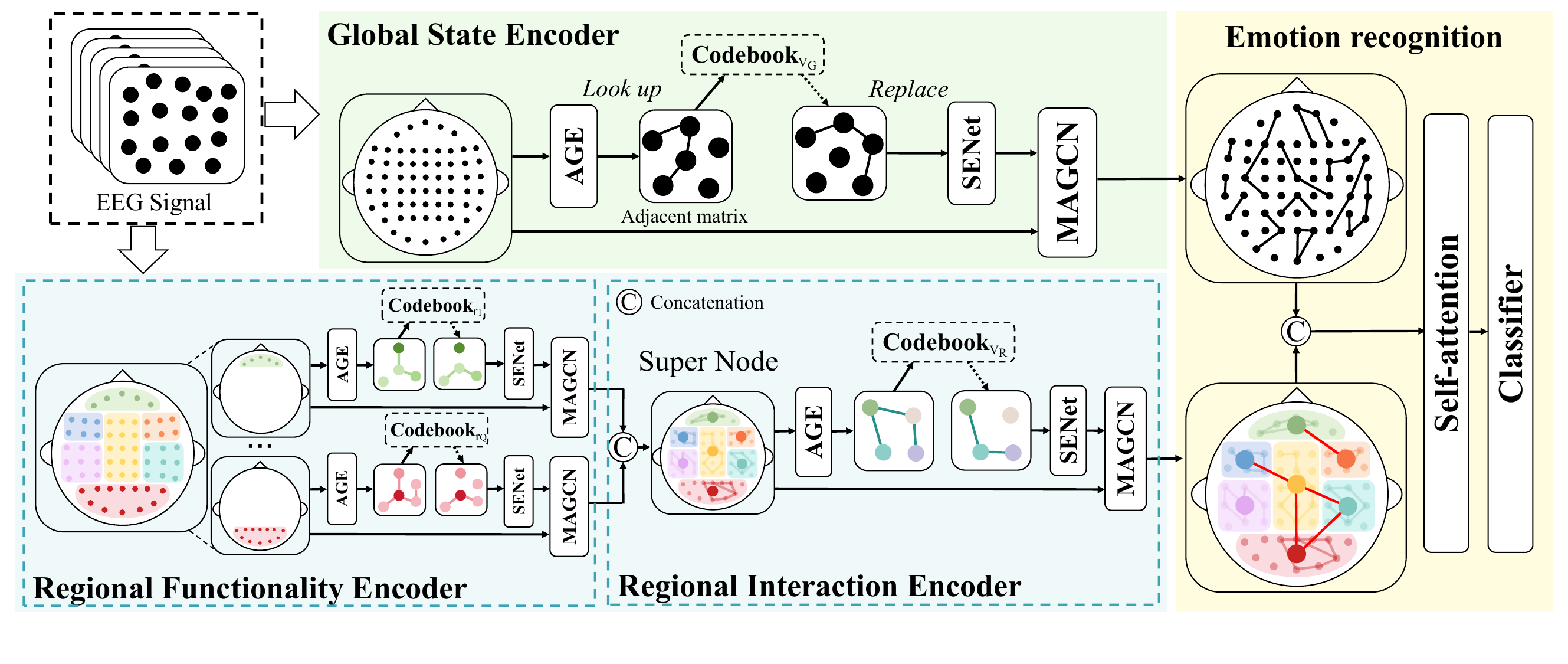}
        \caption{The framework of the proposed MIND-EEG. The Global State Encoder captures the global representation of the entire EEG signal. The Intra-Regional Functionality Encoder and Inter-Regional Interaction Encoder divide the EEG signal into regions, extracting functional features within regions and modeling inter-region interactions. Multi-granularity spatial information is integrated for emotion recognition, with all graph networks in the model constructed using a discrete codebook approach.}
    \label{framework}
\end{figure*}

\section{Related Work}
\subsection{Spatial Structure Modeling}
The spatial information in EEG signals has been modeled in various ways. Early research in EEG signals often modeled the brain as a complete network. For instance, 
Song et al. ~\cite{DGCNN} modeled the brain as a graph network for emotion recognition. 
Some studies have taken into account the asymmetry between the two hemispheres of the brain. 
Ding et al. ~\cite{TSception} proposed a hemispheric convolutional kernel to learn the relationships between the hemispheres. 
%
Recent studies have considered the relationships between different functional regions of the brain~\cite{brainregion}. For example, Li et al. ~\cite{R2G-STNN} used Bidirectional Long Short-Term Memory (BiLSTM) networks to separately learn spatiotemporal EEG features within and between regions. 
However, most existing models are limited to single-scale or simplistic partitioning, failing to capture the brain's multi-layered and multi-granular complexity. 

\subsection{Spatial Connection Representation}
The representation of spatial connections has evolved to address the non-Euclidean nature of EEG nodes. Some methods used a fixed adjacency matrix based on knowledge of neuroscience. For example, Liu et al. ~\cite{SGA-LSTM} introduced an attention vector to select the weights of the EEG nodes, thus extracting more discriminative features. Du et al. ~\cite{MD-GCN} weighted the spatial distance matrix and the relational communication matrix between the EEG nodes to establish the connections. Considering the dynamic nature of brain activity, some research proposed a dynamic adjacency matrix to learn the connections. Song et al. ~\cite{DGCNN} introduced a learnable adjacency matrix to dynamically capture the node relationships. Zhong et al. ~\cite{RGNN} introduced node-wise domain adversarial training and an emotion-aware distribution learning regularized graph network based on learnable adjacency matrices. 

Nevertheless, these dynamic methods construct only a single learned adjacency matrix, without addressing individual differences in EEG node correlations. Song et al. ~\cite{IAG} proposed an instance-adaptive graph model to capture individual-specific relationships between EEG nodes. However, these methods often suffer from overfitting or over-smoothing, particularly with small or augmented datasets, where distinct representations may be compressed into narrow latent spaces. Balancing dynamic network modeling with robustness to overfitting remains a key challenge.

\section{Methodology}
In this section, we introduce the proposed MIND-EEG. We first present the overall framework of MIND-EEG for modeling the spatial structure, followed by a description of the discrete codebook used in representing specific spatial connections. Finally, we detail the integrative loss for emotion recognition and the training procedures.

\subsection{Multi-granularity Integration Network Framework}
In this section, we provide a detailed explanation of how the proposed Multi-granularity Integration Network Framework models spatial information in EEG signals at multiple granularities. As shown in Figure~\ref{framework}, our framework consists of two main parts. The first part, the Global State Encoder, extracts the global representation from the entire EEG signal. The second part, consisting of the Intra-Regional Functionality Encoder and Inter-Regional Interaction Encoder, divides the EEG signal into multiple regions and captures both the functional representations within each region and the inter-region interactions. Below, we describe the processes and model details for each part in turn.

\subsubsection{Global State Encoder}
Following the extraction of energy features from five distinct frequency bands, specifically the $\delta$ band (1–4 Hz), $\theta$ band (4–8 Hz), $\alpha$ band (8–14 Hz), $\beta$ band (14–30 Hz), and $\gamma$ band (30–50 Hz), an EEG sample is expressed as $X \in \mathbb{R}^{n \times d}$. Here, $n$ corresponds to the number of EEG nodes, and $d$ indicates the number of frequency bands. The Global State Encoder evaluates each sample as an integrated whole to extract global state representations. Drawing inspiration from prior works~\cite{IAG,V-IAG}, we dynamically construct a graph network for each input sample by merging spatial and frequency-domain information through our Adapative Graph Encoder (AGE) module. This is achieved by projecting $X$ through a combination of left and right multiplication, formulated as: 
\begin{equation}
   A = \text{ELU}((MX + B)NP), 
   \label{MAP}
\end{equation}
where $M \in \mathbb{R}^{n \times n}$  is the left multiplication matrix used to encode spatial relationships among EEG nodes, $B$  represents the bias matrix, $N \in \mathbb{R}^{d \times d}$ fuses information across frequency bands, and
$P \in \mathbb{R}^{d \times nd}$  is the projection matrix. The resulting output $A \in \mathbb{R}^{n \times nd}$ undergoes activation with the ELU function to ensure non-negativity. Finally, $A$ is reshaped into  $d$ adjacency matrices, $[A^*_1,...,A^*_d]$, corresponding to graphs derived from the $d$ frequency bands.
To normalize the adjacency matrix, each element $A_{ij}$ is scaled by $\frac{1}{\sqrt{D_{ii}D_{jj}}}$. This process is formally expressed as $A^{norm}=D^{-\frac{1}{2}}AD^{-\frac{1}{2}}$, where $D$ is a diagonal matrix with entries calculated as $D_{ii} = \sum_j A_{ij}$.

\begin{figure}
    \centering
        \includegraphics[width=\columnwidth]{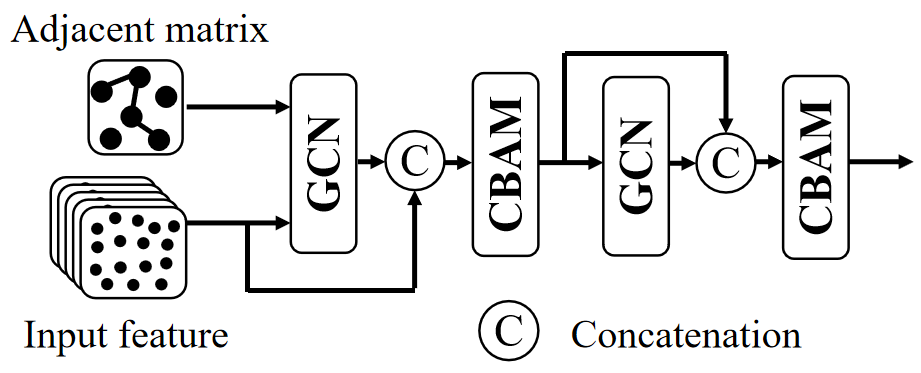}
        \caption{Structure of the MAGCN.}
    \label{MAGCN}
\end{figure}

Once the graph network is constructed for each sample, we apply vector quantization using the codebook to efficiently learn more compact and meaningful latent representations of the brain networ for $d$ frequency. The detailed introduction to the construction of the codebook is provided in Section 3.3. The $d$-frequency brain networks, quantized through the codebook, are then merged using a squeeze-and-excitation network (SENet)~\cite{SE} to produce the global brain network $A_G$ corresponding to the input sample. Subsequently, the brain network and the input sample are fed into a multi-layer attention-based graph convolutional network (MAGCN) module. This module consists of multiple graph convolutional network (GCN) layers, convolutional block attention modules (CBAM)~\cite{CBAM}, and residual layers~\cite{residual}, as shown in Figure~\ref{MAGCN}. 1) The GCN layer computes the Laplacian matrix $\tilde{L}$ associated with the global brain network $A_G$ and facilitates information exchange between adjacent nodes using the following equation:
\begin{equation}
    H^{(l+1)} = \sigma \left( \tilde{L} H^{(l)} W^{(l)} \right),
\end{equation}
where $H^{(l)}$ and $H^{(l+1)}$ denote the node feature matrices at the input and output of the $l$-th layer, respectively. The initial input features $H^{(0)}$ correspond to the original input $X$.  The matrix $W^{(l)}$  is a learnable weight matrix specific to layer $l$, and $\sigma$ is the activation function applied to introduce non-linearity.
2) The CBAM layer, as a general and efficient network, enhances feature representation by adaptively refining the input feature maps through node and spatial attention mechanisms. It is particularly suitable for EEG data feature extraction, allowing the model to focus on the most informative parts of the features.
3) Residual layers help mitigate the vanishing gradient problem by enabling the gradient to flow directly through the network layers. They also improve model convergence and stability by preserving information from earlier layers. The MAGCN module ultimately outputs the extracted global state representations $X_G$.
The SE network and CBAM network, being commonly used plugin modules, are not described in detail here regarding their specific architectures.

\subsubsection{Regional Functionality and Interaction Encoders}
As shown in the lower part of the Figure~\ref{framework}, there are two encoders for regional feature extraction: the Intra-regional Functionality Encoder and the Inter-regional Interaction Encoder, which model the intra-regional and inter-regional spatial information, respectively, and extract the corresponding features. Some components are identical to those in the Global State Encoder and the following focuses on the different parts.

The human cerebral cortex is highly interconnected and centralized across different regions, with some studies dividing it into multiple emotion-related brain regions~\cite{hagmann2008mapping,bruder2017right}. For different tasks and scenarios, we can partition the EEG data into varying numbers of regions and input them into our model. Therefore, the input EEG sample $X \in \mathbb{R}^{n \times d}$  is divided into $\{X_1,...,X_Q\}$, where $X_i \in \mathbb{R}^{n_i \times d}$, $Q$ is the number of the regions and $\sum_i n_i = n$ ($i=1,...,Q$).  
Similar to the Global State Encoder, each regional sample is used to obtain the corresponding initial intra-regional brain network through Eq.~(\ref{MAP}). After being quantized by their respective codebooks, the networks are input into the SE module and MAGCN module to extract the intra-regional EEG features 
$X_R=\{X_{R_1},...,X_{R_Q}\}$. Before extracting inter-regional features, we need to fuse the intra-regional features from all regions. Specifically, for each $X_{R_{i}}\in \mathbb{R}^{n_i \times d}$, the attention-based connectivity matrix $a$ is calculated as follows:
\begin{equation}
    a = X_{R_{i}} W{(X_{R_{i}} W)}^T,
\end{equation}
where $W$ is a trainable weight matrix, and $\cdot^T$ represents the transpose operation. Subsequently, the weight coefficient $C_{i} \in \mathbb{R}^{1 \times n_i}$ for each region is obtained by performing a row-wise summation on $a$. Here, $C_i$ reflects the functional connection strength of each node relative to all other nodes within the region. Finally, the fused regional features are computed as:
\begin{equation}
    \hat{X_{R_{i}}} = softmax(C_{i}) X_{R_{i}}.
\end{equation}

In the Inter-regional Interaction Encoder, the fused intra-regional features $\hat{X_R} = \{\hat{X_{R_{i}}}, i=1,..,Q\}$ are used to obtain the corresponding initial inter-regional brain network through Eq.~(\ref{MAP}). The inter-regional brain network is then quantized using the inter-regional graphical codebook and input into the SE module and MAGCN module to extract the final regional EEG features $\overline{X_R}$.

\begin{figure}
    \centering
        \includegraphics[width=\columnwidth]{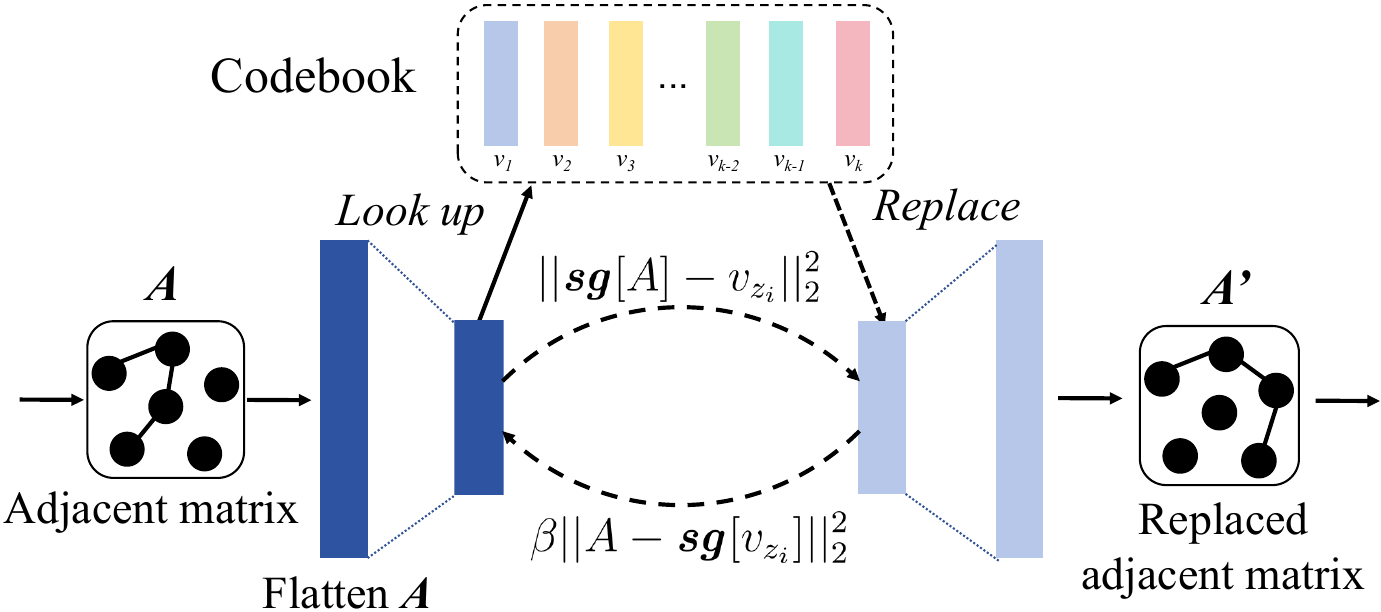}
        \caption{Discerte Codebook Construction.}
    \label{codebook}
\end{figure}

\subsection{Discerte Codebook Construction}
This section provides a detailed explanation of the codebook mechanism within the model, as shown in the Figure~\ref{codebook}. The discerte codebook stores discrete representations for vector quantization of the dynamically generated network representations for each sample. It assigns each sample a compact, versatile brain network representation from the codebook, preventing similar samples from converging to overly similar embeddings and mitigating over-smoothing. Additionally, discrete embeddings capture inter-class differences, maintaining clear distinctions between categories and improving classification and generalization.
We construct codebook for the global brain network denoted as $V_G$, $Q$ codebooks for intra-regional brain networks denoted as $V_{r_1}$ to $V_{r_Q}$, and codebook for the inter-regional brain network denoted as $V_R$. The construction and training methods are similar. The global brain network codebook is used as an example below.

Specifically, We define a global graphical codebook $V_G = \{v_i|i=1,...,K\}\in \mathbb{R}^{K \times D}$, where $K$ is the number of the embeddings in codebook and $D$ is the dimensionality of each embedding. The embeddings contained in the codebook represent different brain network states.
To reduce computational cost and enable the codebook to learn more compact representations, the brain network $A$ obtained from the input sample $X$ is flattened and then passed through a linear layer for dimensionality reduction, resulting in $\hat{A}$. Then, the global graphical codebook identifies the nearest neighbor of $\hat{A}$. This operation is expressed as:
\begin{equation}
    z_i = \underset{j}{{\arg\min} } \, ||\hat{A}-v_i||_{2}
\end{equation}
where  $z_i$  represents the quantized vector obtained from the codebook. This process is equivalent to finding the embedding with the highest similarity to the input embedding, typically measured by cosine similarity. 
The output $z_i$ is transformed back to its original dimensions and converted into the adjacency matrix used in subsequent processes.
To update both the network and the embeddings in the codebook, we use the following loss function:
\begin{equation}
    \mathcal{L}_{BG} = ||\textbf{\textit{sg}}[\hat{A}]-v_{z_i}||^{2}_{2} + 
    \beta ||\hat{A}-\textbf{\textit{sg}}[v_{z_i}]||^{2}_{2}
\end{equation}
where $\textbf{\textit{sg}}$ denotes the stop-gradient operation. This function acts as an identity during the forward pass but blocks gradient flow during backpropagation. The first term minimizes the distance between the graph $\hat{A}$ and its corresponding embedding $v_{z_i}$ , while the second term encourages the codebook to adapt to the network's representations. $\beta$ balances the weights of the two parts of the loss and is empirically set to 0.25~\cite{VQ-VAE}. Similarly, the intra-regional graphical codebook and inter-regional graphical codebook are updated using the same loss function $\mathcal{L}_{BR1}$ and $\mathcal{L}_{BR2}$.

\subsection{Integrative Loss and Training}
After obtaining the global and local EEG features, we concatenate them and input the result into an attention layer, followed by the final fully connected classification layer, to obtain the final emotion recognition results. We employ the cross-entropy loss function to measure the dissimilarity between the predicted labels and the real labels, with the classification loss denoted as $\mathcal{L}_C$. The overall loss function of our framework is summarize as:
\begin{equation}
    \mathcal{L} = \mathcal{L}_{C} + \alpha\mathcal{L}_{BG} + \beta\mathcal{L}_{BR1} + \gamma\mathcal{L}_{BR2},
\end{equation}
where $\alpha$, $\beta$, and $\gamma$ are hyper-parameters to balance the classification loss and the losses from the three codebooks.

This paper uses PyTorch to build MIND-EEG and deploy it on a 3090 GPU. The global module contains map layer and MAGCN layer with a size of 62 $\times$ 5 for the input and 62 $\times$ 50 for the output. 
For the model's functional region division, we follow the excellent previous work~\cite{PGCN} and set the number of regions $Q=7$.
Thus, the Intra-regional module contains 7 parallel MAGCN layers with an output size of 78 $\times$ 50. The Inter-regional module contains fusion layer and MAGCN layer with an output size of 7 $\times$ 60; the emotion recognition network contains a 3-layer fully connected network with an input of 69 $\times$ 145. For the optimization of the model, we used the SGD optimizer , setting the learning rate to 1e-2 and the batch size to 32. We evaluated the model using the average accuracy (ACC) and standard deviation (STD).

\section{Experiments}

\begin{table*}[t]
	\caption{The Mean Accuracies and Standard Deviations of Proposed MIND-EEG and Comparison Methods for Subject-Dependent and Subject-Independent EER Experiments on the SEED-IV, MPED, and SEED-V Datasets.}
	\label{dependent}
	\renewcommand{\arraystretch}{1.3}
	\centering
        \begin{tabular}{m{5cm}<{\centering}m{2cm}<{\centering}m{2cm}<{\centering}m{2cm}<{\centering}m{2cm}<{\centering}m{2cm}<{\centering}}
        \toprule
        \multirow{2}*{\textbf{Method}}&\multicolumn{3}{c}{\textbf{Subject-dependent}} &\multicolumn{2}{c}{\textbf{Subject-independent}} \\
        \cmidrule(lr){2-4}\cmidrule(lr){5-6}
        ~&\textbf{SEED-IV} & \textbf{MPED} & \textbf{SEED-V}&\textbf{SEED-IV} & \textbf{MPED}\\
        \midrule
        \textbf{DBN}~\cite{DBN}&66.77$\pm$07.38&29.26$\pm$09.19&----&----&----\\
        \textbf{SVM}~\cite{SVM}&56.61$\pm$20.05&31.14$\pm$08.06&69.5$\pm$10.28&37.99$\pm$12.52&19.66$\pm$03.96\\
        \textbf{GSCCA}~\cite{GSCCA}&69.08$\pm$16.66&----&----&----&----\\
                    \midrule
        \textbf{BiDANN-S}~\cite{BiDANN-S}&70.29$\pm$12.63&40.34$\pm$07.59&----&65.59$\pm$10.39&----\\
        \textbf{A-LSTM}~\cite{MPED}&69.50$\pm$15.65&38.74$\pm$07.75&----&55.03$\pm$09.28&24.06$\pm$04.58\\
        \textbf{BiHDM}~\cite{BiHDM}&74.35$\pm$14.09&40.34$\pm$07.59&----&69.03$\pm$08.66&25.17$\pm$04.99\\
        \textbf{EeT}~\cite{EeT}&83.27$\pm$08.37&----&----&----&----\\
        \textbf{DMMR}~\cite{DMMR}&----&----&----&72.70$\pm$08.01&----\\
        \midrule
        \textbf{DGCNN}~\cite{DGCNN}&69.88$\pm$16.29&36.92$\pm$12.78&----&52.82$\pm$09.23&25.12$\pm$04.20\\
        \textbf{RGNN}~\cite{RGNN}&79.37$\pm$10.54&----&----&73.84$\pm$08.02&----\\
        \textbf{IAG}~\cite{IAG}&----&40.38$\pm$08.75&----&62.64$\pm$10.25&27.11$\pm$05.55\\
        \textbf{MD-AGCN}~\cite{MD-AGCN}&87.63$\pm$05.77&----&80.77$\pm$06.61&----&----\\
        \textbf{V-IAG}~\cite{V-IAG}&----&40.40$\pm$09.35&----&----&----\\
        \textbf{Siam-GCAN}~\cite{Siam-GCAN}&79.45$\pm$15.14&----&----&----&----\\
        \textbf{GMSS}~\cite{GMSS}&86.52$\pm$06.22&40.16$\pm$06.08&----&73.48$\pm$07.41&\underline{28.49$\pm$04.42}\\
        \textbf{MSFR-GCN}~\cite{MSFR-GCN}&\underline{89.02$\pm$11.31}&----&----&73.43$\pm$07.32&----\\
        \textbf{PGCN}~\cite{PGCN}&82.24$\pm$14.85&\underline{40.54$\pm$07.02}&81.67$\pm$08.39&73.69$\pm$07.16&----\\
        \textbf{BGAGCN-MT}~\cite{BGAGCN-MT}&82.86$\pm$09.31&----&----&\textbf{75.78$\pm$08.17}&----\\
        \textbf{SGGT}~\cite{SGGT}&88.62$\pm$08.01&----&\underline{82.53$\pm$05.84}&----&----\\
        \midrule
        \textbf{MIND-EEG}&\textbf{92.21$\pm$09.05}&\textbf{42.08$\pm$06.11}&\textbf{85.03$\pm$09.93}&\underline{74.79$\pm$08.66}&\textbf{29.50$\pm$04.75}\\
        \bottomrule
        \end{tabular}

\end{table*}
\subsection{Experimental Settings}
\subsubsection{Datasets}
The SEED-IV~\cite{SEED-IV}, SEED-V~\cite{SEED-V}, and MPED~\cite{MPED} datasets all include EEG recordings collected while participants watched movie clips designed to evoke various emotional states. The SEED-IV dataset consists of EEG data from 15 participants (7 males and 8 females) across three sessions, with each session containing 24 trials, corresponding to 2-minute movie clips inducing four emotional states: neutral, sad, fear, and happy. Similarly, the SEED-V dataset includes EEG recordings from 16 participants (10 females and 6 males) over three sessions. In total, 45 movie clips were shown to evoke five emotional states: happy, disgust, neutral, fear, and sad. The MPED dataset, which includes data from 23 participants (10 males and 13 females), consists of EEG recordings for 28 movie clips designed to trigger seven emotions: joy, funny, anger, fear, disgust, sadness, and neutrality. For all datasets, EEG signals were segmented into 1-second samples.

\subsubsection{Comparison Methods}
We select a wide range of representative and well-established models in this field as comparison methods. Some of these models are non-graph-based approaches, including three representative algorithms based on machine learning 
and five representative non-graph-based deep learning models. 
In addition, we compare our method with a variety of graph-based deep learning models, including traditional classic models and the latest state-of-the-art (SOTA) approaches. The list of comparison methods is in Table~\ref{dependent}.

\subsubsection{Expreimental Protocol}
We evaluated the proposed PGCN using both subject-dependent and subject-independent protocols.
In subject-dependent experiments, training and test data came from the same subject, with average accuracy and standard deviation computed across all subjects. For SEED-IV, we followed~\cite{RGNN}, using the last two trials of each emotion for testing and the remaining 16 for training. For SEED-V, we adopted the three-fold cross-validation setup in~\cite{MD-AGCN}, splitting each session into first, middle, and last five trials. For MPED, as in~\cite{MPED}, the first 21 trials were used for training and the last 7 for testing.
In subject-independent experiments, training and test data were from different subjects. For SEED-IV, we applied leave-one-subject-out cross-validation per~\cite{BiHDM}, averaging results and standard deviations across subjects. For MPED, we used the same leave-one-subject-out approach as~\cite{MPED}. As no comparison methods conducted subject-independent experiments on SEED-V, we excluded it from such experiments.

All datasets underwent identical preprocessing to ensure fairness. SEED-IV and SEED-V used differential entropy (DE) features~\cite{DE}, while MPED employed short-time Fourier transform (STFT) features~\cite{STFT}.

\begin{table}[t]
	\caption{The results of ablation experiments.}
	\label{ablation}
	\renewcommand{\arraystretch}{1.3}
	\begin{center}
	{\normalsize{
            \begin{tabular}{m{2.7cm}<{\centering}m{2cm}<{\centering}m{2cm}<{\centering}}
            \toprule
            \multirow{2}*{\textbf{Method}} & \multicolumn{2}{c}{\textbf{SEED-IV}} \\
            \cmidrule(lr){2-3} 
            ~ &ACC(\%)&F1(\%) \\
            \midrule
            Regional-removed  & 80.82$\pm$12.39 &  76.10$\pm$14.82\\
            Global-removed   & 85.39$\pm$12.22&81.49$\pm$15.18 \\
            Intra-removed  & 84.80$\pm$11.68&81.30$\pm$15.17\\
            Inter-removed   & 86.16$\pm$09.71&83.88$\pm$11.81 \\
             \midrule
            MIND-EEG  &\textbf{92.21$\pm$09.05}&\textbf{90.04$\pm$11.06} \\
            \bottomrule
            \end{tabular}
	}}
    \\
    \end{center}
    \vspace{-0.2cm}
\end{table}

\subsection{Experimental Results and Analyses}
\subsubsection{Overall Comparison}
In this section, we conduct experiments in the subject-dependent and subject-independent scenarios. Table~\ref{dependent} reports the average accuracies and standard deviations of the proposed MIND-EEG and all comparison methods on the SEED-IV, MPED, and SEED-V datasets. The results of all comparison methods in the table are consistent with those reported in the respective published papers. Those ``----" in the table indicates that the corresponding paper did not conduct experiments on the respective dataset.

As shown in the Table~\ref{dependent}, the proposed MIND-EEG achieves the outstanding average accuracy on all three datasets compared to all the comparison methods under both scenarios, while maintaining comparable standard deviations. Specifically, in the subject-dependent task, our model improves the accuracy by 3.19\%, 1.54\%, and 2.50\% over the best-performing comparison methods on the SEED-IV, MPED, and SEED-V datasets, respectively. In the subject-independent task, our model also achieved competitive results, demonstrating strong generalization capabilities. This result fully demonstrates our model's superiority.
\begin{figure} [!t]
	
	\begin{center}    
		\subfloat[] {    
			\includegraphics[width=0.45\columnwidth]{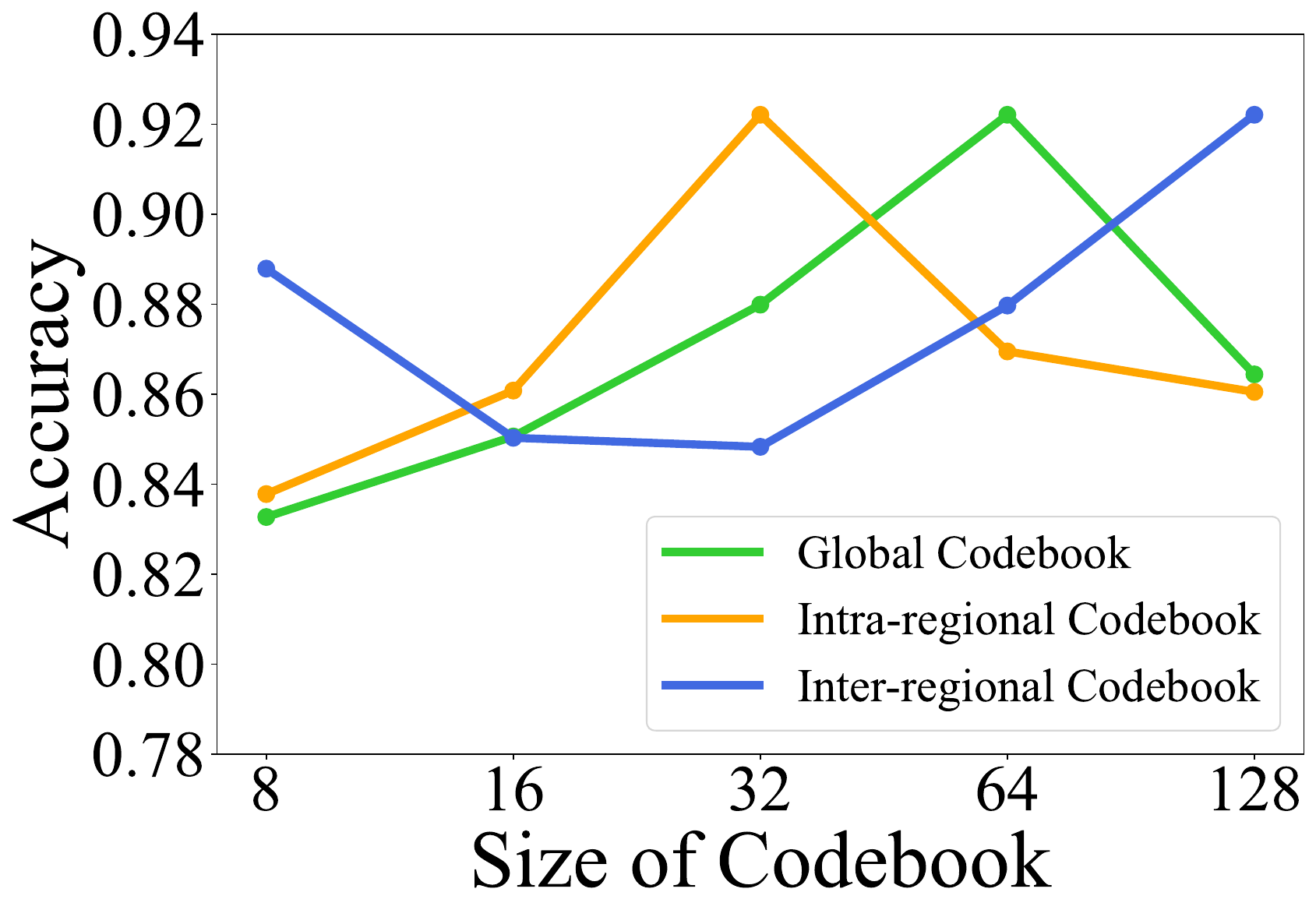}  
		}     
		\subfloat[] {     
			\includegraphics[width=0.45\columnwidth]{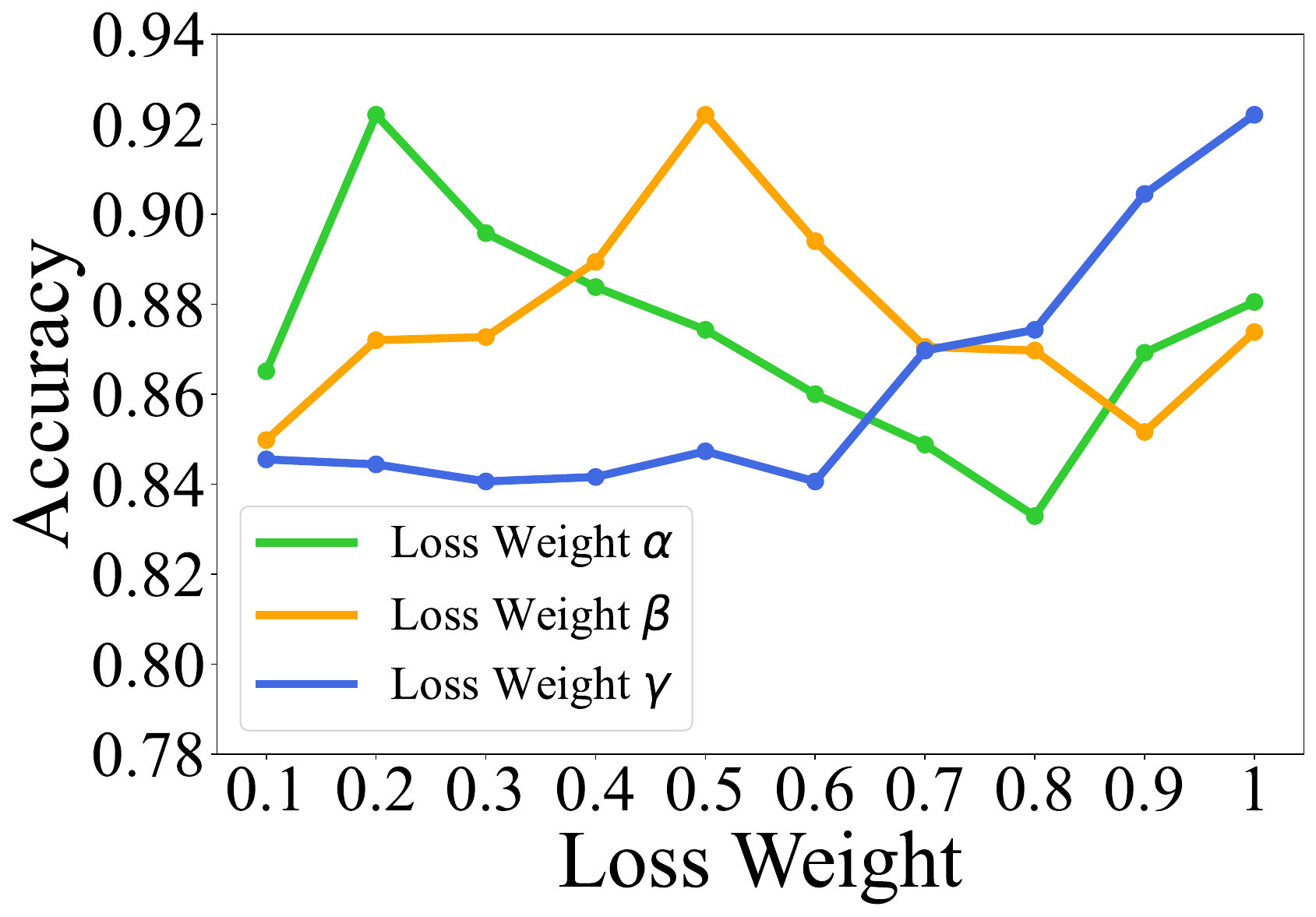}     
		}   

	\end{center}
	\caption{The parameter analysis of our proposed method on the SEED-IV dataset with subject-dependent scene. (a) Analysis of the sizes of three types of codebooks. (b) Analysis of the weight of the three codebook loss terms.}
	\label{fig:parameter}  
\end{figure}

\subsubsection{Ablation Studies}
To thoroughly evaluate the impact of each module in our model, we conducted ablation studies on the three datasets with ACC and F1 score in the subject-dependent scene. In these studies, we systematically removed individual modules (the whole regional part, the global part, the intra-regional part, and the inter-regional part) from our model to demonstrate the effectiveness of each module. The results of the ablation experiments on the SEED-IV dataset are shown in Table~\ref{ablation}, while the results on the other two datasets are provided in the appendix due to space limitations. From these results, we can draw the following findings: 1) Removing any module leads to a decline in model performance, demonstrating that each part plays a crucial role and that they complement each other to achieve the best performance.
2) Compared to removing the global part, removing the regional part results in a more significant performance drop. This suggests that the regional module (both intra-regional and inter-regional) are more important. Although the global part is also crucial, multi-layer and finer-grained graph modeling contributes more to feature extraction. 3) The results of removing either the intra-regional or inter-regional components show that both modules are meaningful, with the intra-regional module being somewhat more important.

\subsubsection{Hyperparameter Analysis}
In this section, we explore the selection of two critical groups of hyperparameters in the model. The first group involves the sizes of the codebooks in the three modules: the global codebook, intra-regional codebook, and inter-regional codebook, denoted as $K_1$, $K_2$, and $K_3$, respectively. For this analysis, we vary one parameter from $\{8,16,32,64,128\}$ while keeping the other two fixed. 
The second group consists of the weights for the three codebook loss terms: $\alpha$, $\beta$, and $\gamma$. Similarly, we vary one weight within the range $\{0.1,0.2,0.3,0.4,0.5,0.6,0.7,0.8,0.9,1\}$ while keeping the other two constants. The results of these experiments, conducted on the SEED-IV dataset under the subject-dependent scenario, are presented in Figure~\ref{fig:parameter}. Based on the results in Fig.~\ref{fig:parameter} (a) and (b), we set $K_1$ to 32, $K_2$ to 64, $K_3$ to 128, $\alpha$ to 0.2, $\beta$ to 0.5, and $\gamma$ to 1. The results indicate that the model's performance is highly sensitive to the two groups of parameters related to the crucial codebook mechanism.
 \begin{figure}
    \centering
    \subfloat[MPED dataset] {    
			\includegraphics[width=0.8\columnwidth]{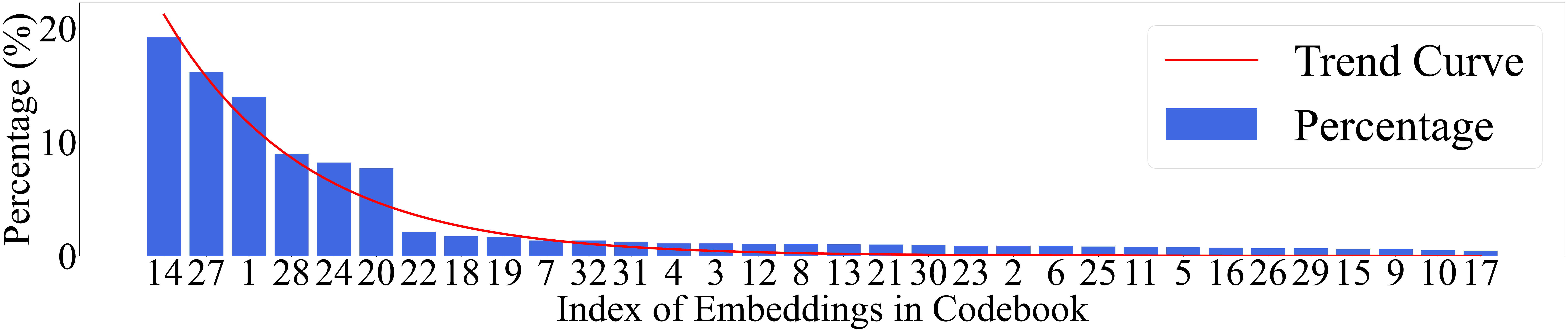}  
		} \\
    \subfloat[SEED-IV dataset] {    
			\includegraphics[width=0.8\columnwidth]{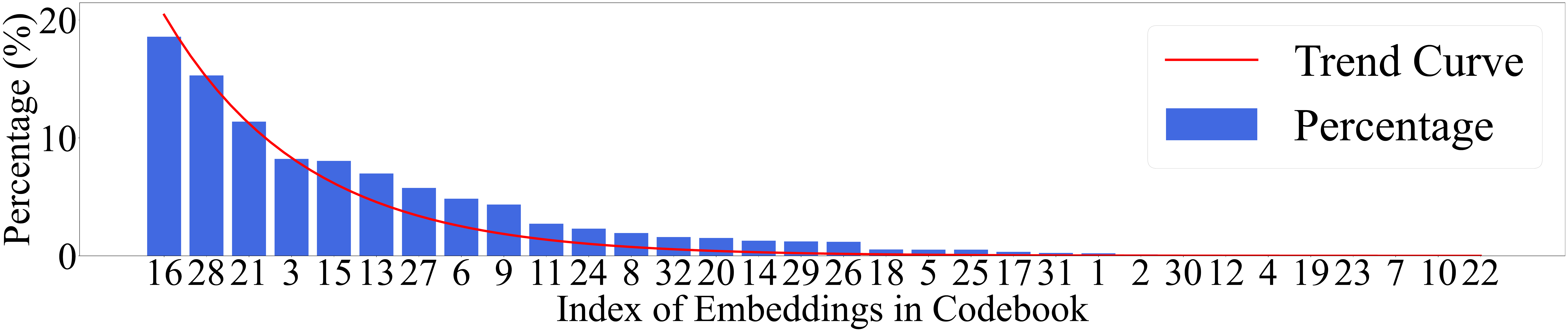}  
		} \\
    \subfloat[SEED-V dataset] {    
			\includegraphics[width=0.8\columnwidth]{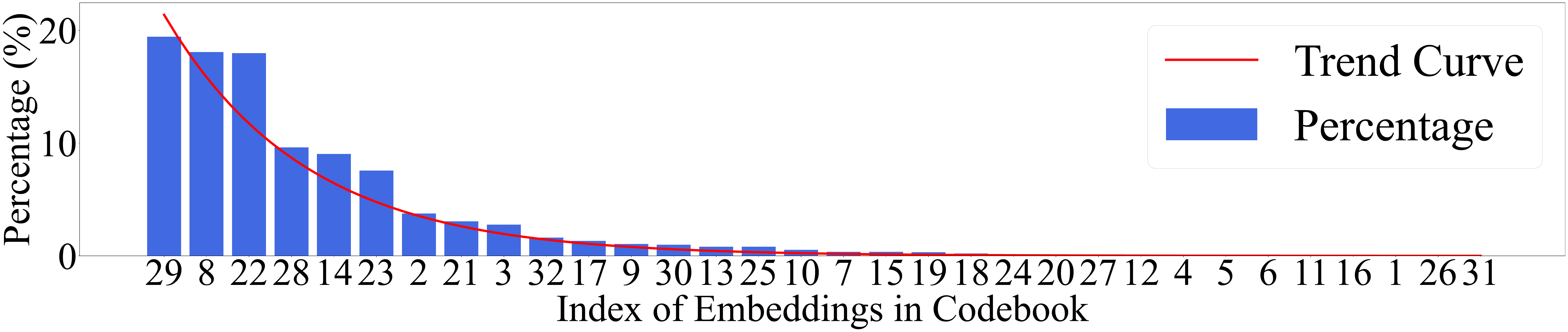}  
		}

        \caption{The usage of embeddings in the global codebook across the three datasets.}
    \label{code_analy}
    \vspace{-0.2cm}
\end{figure}

\subsubsection{Effectiveness of Codebook}
In this section, we discuss the effectiveness of the codebook. We conducted a statistical analysis on the usage proportion of embeddings in the codebook for all samples from the first subject across the three datasets. The x-axis represents embedding indices in the codebook, sorted by usage frequency. The y-axis shows the percentage of samples using each embedding. Due to space constraints, we only present the results of the global codebook in Fig~\ref{code_analy}, while the complete results for all codebooks are included in the appendix. From the results, we observe that for each dataset, a large number of embeddings are utilized. This indicates that the codebook has generated multiple distinct network structure representations for different samples. Additionally, in each case, some embeddings are used extensively, which we believe represent common structural features in the EEG signals, while others are used sparingly, representing more individualized characteristics of the network structure.
In future work, we plan to further explore the interpretability of these embeddings in the codebook.

\section{Conclusion and Future Work}
This work presents the Multi-Granularity Integration Network with Discrete Codebook (MIND-EEG), a novel framework for EEG-based emotion recognition. By integrating spatial information across multiple levels and utilizing a discrete codebook mechanism, the proposed model effectively models the dynamic and diverse characteristics of brain networks while addressing challenges such as overfitting and over-smoothing.
Extensive experiments on three benchmark datasets demonstrate the superiority of MIND-EEG over existing methods in both subject-dependent and subject-independent scenarios, with significant accuracy improvements. Analysis of the differences and utilization of network structure representations within the codebook also validates the effectiveness of this design. Future work will focus on further exploring the relationship between brain network connectivity and emotional states using the codebook to enhance model interpretability.


\bibliographystyle{named}
\bibliography{ijcai25}

\end{document}